\documentclass{aa}

\include{psfig}

\def\e20{\times 10^{20} {\rm cm}^{-2} }
\def\sax{{\it Beppo}SAX}

\def\ecs{\rm \,erg~ cm^{-2}\,s^{-1} }
\def\mincir{\ \raise -2.truept\hbox{\rlap{\hbox{$\sim$}}\raise5.truept  
\hbox{$<$}\ }}                        

\begin{document}

\thesaurus{03(11.02.1; 13.25.3)}	

\title{X--ray variability and prediction of TeV emission in the HBL 1ES~1101--232}
\author{A. Wolter,
\inst{1} 
F. Tavecchio, \inst{1} A. Caccianiga,\inst{2} G. Ghisellini,\inst{1} G. Tagliaferri \inst{1}
}

\offprints{A. Wolter: anna@brera.mi.astro.it}

\institute{Osservatorio Astronomico di Brera,
		Via Brera, 28
		20121 MILANO, Italy
\and Observatorio Astronomico de Lisboa, Tapada da Ajuda, P-1300 Lisboa, Portugal
}

\date{Received ....; accepted ....}
\maketitle
\markboth{A. Wolter et al.}{The HBL 1ES~1101--232}{}

\begin{abstract}

1ES~1101--232 is a bright BL Lac of the High frequency peak class. 
We present here the results of two \sax\ observations in which the
source has shown a variation of about 30\% in flux with a corresponding
spectral variability.
We interpret the overall spectral energy distribution in terms
of an homogeneous SSC model and, by using also the TeV upper limit
from a short Mark 6 pointing, derive constraints
on the physical parameters of the source, in particular on the magnetic 
field strength.
The overall Spectral Energy Distribution makes
1ES~1101--232 a very promising candidate for TeV detection. 

\keywords{(Galaxies:) BL Lacertae objects: general --
	  X--rays: galaxies -- BL Lacertae objects: individual: 1ES~1101--232
         }
\end{abstract}

\section{Introduction}
\noindent

BL Lac objects form a minority class of active nuclei (see e.g. Urry \& 
Padovani, 1995), but nevertheless their high luminosities and extreme 
variability in all bands 
make them an interesting subclass to study intrinsic properties of
nuclear emission. Furthermore, it is matter of discussion the position of
BL Lacs with respect to other active nuclei in the framework of Unification
Models. The current picture claims that BL Lacs are the fraction of 
Fanaroff-Riley I galaxies that point their jet towards us, but many
details still need to be worked out. Often, BL Lacs are studied together
with other classes of flat radio spectrum sources, to form the class of blazars.

Recently a sequence has been proposed for the subclass of gamma ray
bright blazars, and possibly valid for all blazars, based on their 
bolometric luminosity
(Ghisellini et al. 1998, Fossati et al. 1998). 
The overall spectral energy distribution has two peaks 
(in $\nu F_\nu$ representation), 
the one at lower energies due to synchrotron radiation and the higher energy 
one to Inverse Compton scattering. 
It is proposed that the value of the peak frequency is the result
of the balance between radiative cooling and acceleration of the 
corresponding electrons (see e.g. Ghisellini, 1999):
in less luminous objects the radiative cooling is less efficient, allowing
the accelerated electrons to reach higher energies.
As a consequence, both the synchrotron and the inverse Compton peaks
shift to higher frequencies as the bolometric intrinsic
luminosity decreases.
In this view, the class of BL Lacertae objects known as HBL 
(High frequency peak BL Lacs)
has a smaller bolometric luminosity and a higher synchrotron peak frequency
than the class of LBL (Low frequency peak BL Lacs).
HBL are mostly found in X--ray selected samples, since they are expected to 
produce most of their synchrotron emission in the X--ray band. 
If the emission peak (as in the LBL objects) is
at frequencies smaller than the observed X--ray band, the X--ray spectrum is 
steep (i.e. the steepening part above the peak), while HBL with a flat 
X--ray spectrum should have their peak in the observed X--ray band.

As part of a program aimed at a spectral survey of soft X--ray selected 
BL Lacs with \sax\ (Wolter et al. 1998), 
we have studied also 1ES~1101--232 (z=0.186), 
a bright BL Lac selected from the Slew Survey (Perlman et al. 1996) 
that shows an extreme behavior with a very flat X--ray spectrum.
The object was detected, besides in the 0.1-10 keV energy band, also
up to $\sim$ 100 keV in only $\sim 6000$ sec. 
The best fit of the source in the 0.1-100 keV band is given by a broken 
power law, with Galactic low energy absorption, that has a break energy
E$_0$=1.36 (1.11--1.65) keV.
The spectral slope (energy index) derived from the PDS is consistent
with the one derived from the MECS above the break energy E$_0$:
$\alpha_x$ = 1.03 (0.99--1.08) (Wolter et al. 1998).

We have constructed also the Spectral Energy Distribution (SED) for this
object, using flux measurements collected from the literature, from radio to
X--rays.
We have fitted a cubic polynomial to the 
distribution in order to find the peak of the SED, that indeed 
falls in the \sax\ band (log $\nu_{peak} \sim$ 17.48, corresponding to 
$\sim 1.3$ keV).
For this object, therefore, the break energy derived from
the spectral fit in the X--ray band is consistent, albeit within its
large indetermination, with the position of the synchrotron 
peak as derived from the overall distribution (SED). 

The SED of 1ES~1101--232 is similar to that of the flaring state
of Mkn 501 (Pian et al. 1998) and 1ES~2344+514 (Catanese et al. 1998; 
Giommi et al. 2000), 
and therefore we could expect a strong TeV emission. 
A quasi--simultaneous X--ray and TeV observation has been
therefore scheduled, to confirm the X--ray spectrum of the source, and
its overall shape,
to detect possible variations in flux, that could constrain the physical
parameters of the source, and to monitor the TeV emission to search
for possible detection. 

\begin{table*}
 \caption{Fit results for 1ES~1101--232 LECS+MECS data OBS. 19/06/98 {\bf Low state}}
\begin{tabular}[h]{| l l c c c r r r|}
            \hline
Model  &   $N_{\rm H}^a$ & $\alpha_1$ & $\alpha_2$   & $E_0$     & $F^{\rm b}$ &  $\chi^2(dof)$ & Prob. \\
       &           &               &                 & keV       &             &  & \\
(1)  & 5.76        &    --         & 1.19(1.16-1.22) & --        & 2.60     & 267.1(195) & $<$ 0.5\%\\
(2)  & 8.3(7.1-10.4) &  --         & 1.25(1.22-1.29) & --        & 2.57     & 240.5(195) & 1.5\%\\
(3)  & 5.76        &0.80(0.58-0.96)& 1.29(1.25-1.33) & 1.34(1.08-1.82) &2.55 & 216.2(194) & 13\% \\
\hline
\end{tabular}
\begin{list}{}{}
\item (1) Single p.l., $N_{\rm H}$=$N_{\rm H}^{\rm Gal}$, LECS/MECS ratio free; 
(2) Single p.l., $N_{\rm H}$ free, LECS/MECS ratio=0.7 ; 
(3) Broken p.l., $N_{\rm H}=N_{\rm H}^{\rm Gal}$, LECS/MECS ratio = 0.7. 
Errors quoted are 90\% confidence intervals. \\
$^{\rm a}$ Column density in $\e20$. \\
$^{\rm b}$ Unabsorbed [2--10 keV] flux in $10^{-11} \ecs$. \\
\end{list}
\label{broken_n}
\end{table*}
\begin{table*}
 \caption{Fit results for 1ES~1101--232 LECS+MECS data OBS. 04/01/97 {\bf High state} } 
\begin{tabular}[h]{| l l c c c r r r|}
            \hline
Model  &   $N_{\rm H}^{\rm a}$ & $\alpha_1$ & $\alpha_2$ & $E_0$  & $F^{\rm b}$ &  $\chi^2(dof)$ & Prob. \\
       &           &               &                 & keV        &             &  & \\
(1)  & 5.76          &  --          & 0.97(0.95-1.00) & --        & 3.81       & 224.0 (182) & $<$ 2\%\\
(2)  & 8.9(7.2-12.6) &  --          & 1.03(0.99-1.08) & --        & 3.79       & 205.1 (182) &  11 \%\\
(3)  & 5.76      &0.59(0.35-0.74)   & 1.05(1.01-1.08) & 1.36(1.11-1.65) &3.76  & 191.6 (181) & 28 \%\\
\hline
\end{tabular}
\begin{list}{}{}
\item 
(1) Single p.l., $N_{\rm H}$=$N_{\rm H}^{\rm Gal}$, LECS/MECS ratio free; 
(2) Single p.l., $N_{\rm H}$ free, LECS/MECS ratio=0.7; 
(3) Broken p.l., $N_{\rm H}$=$N_{\rm H}^{\rm Gal}$, LECS/MECS ratio = 0.7.
Errors quoted are 90\% confidence intervals. \\
$^{\rm a}$ Column density in $\e20$. \\
$^{\rm b}$ Unabsorbed [2-10 keV] flux in $10^{-11} \ecs$. \\
\end{list}
 \label{broken_o}
\end{table*}

The \sax\ data are presented here. The TeV observation, conducted in 
non--optimal weather condition, did not yield a detection, but we will use
the upper limit (Chadwick et al. 1999a) to derive useful information on
the physical mechanisms at work in this source.
The plan of the paper is as follows: in Section 2 we describe the observational
data obtained with \sax\ in the two epochs, in Section 3 we summarize the TeV
predictions and observations, that help constraining the parameters of the
source, by using the SED and theoretical models of emission as explained in
Section 4. Section 5 presents our results and conclusions.

Throughout the paper a Hubble constant H$_0$=50 $km\, s^{-1}\, Mpc^{-1}$ 
and a Friedman universe with a deceleration parameter $q_0$=0 are assumed.

\section{ \sax\ data}

The X--ray astronomy satellite \sax\ is a project of the Italian
Space Agency (ASI) with a participation of the Netherlands Agency for 
Aerospace Programs (NIVR).
The scientific payload comprises   four Narrow Field Instruments [NFI:
Low Energy Concentrator Spectrometer (LECS), Medium Energy Concentrator
Spectrometer (MECS), High Pressure Gas Scintillation Proportional Counter
(HPGSPC), and Phoswich Detector System (PDS)], all pointing in the
same direction, and two Wide Field Cameras (WFC), pointing in
opposite directions perpendicular to the NFI common axis. A detailed
description of the entire \sax\ mission can be found in Butler \& Scarsi (1990)
and Boella et al. (1997a). 

The MECS consists of three equal units, each composed of a grazing
incidence mirror unit and of a position sensitive gas scintillation 
proportional counter, with a field of view of
56 arcmin diameter, working range 1.3--10 keV, energy resolution $\sim 8\%$
and angular resolution $\sim 0.7$ arcmin (FWHM) at 6 keV.
The effective area at 6 keV is 155 cm$^2$ (Boella et al. 1997b)

The LECS is a unit similar to the MECS, with a thinner window that
grants a lower energy cut-off (sensitive in the energy range 0.1-10.0 keV)
but also reduces the FOV to 37 arcmin diameter (Parmar et al. 1997). 
The LECS energy resolution is a factor $\sim 2.4$ better
than that of the ROSAT PSPC ($\sim 32\%$ at 0.28 keV), while the 
effective area is smaller: 22 cm$^{2}$ at 0.28 and and 50 cm$^{2}$ at 6 keV.

The PDS is a system of four crystals, sensitive in the 13--200~keV band and
mounted on a couple of rocking collimators, which points two units on the
targets and two units $3.5^{\circ}$ aside respectively, to monitor the
background.
The position of the collimators flips every 96 seconds.
Thanks to the stability of the instrumental background, the PDS has shown 
an unprecedented sensitivity in its energy range, allowing 3$\sigma$ detection 
of $\alpha \sim 1$ sources as faint as 10 m Crab 
with 10 ks of effective exposure time (Guainazzi \& Matteuzzi, 1997).

The source is not detected by the HPGSPC, so we will not discuss this 
instrument.

\subsection{Observation of June 1998}

\begin{figure}
\psfig{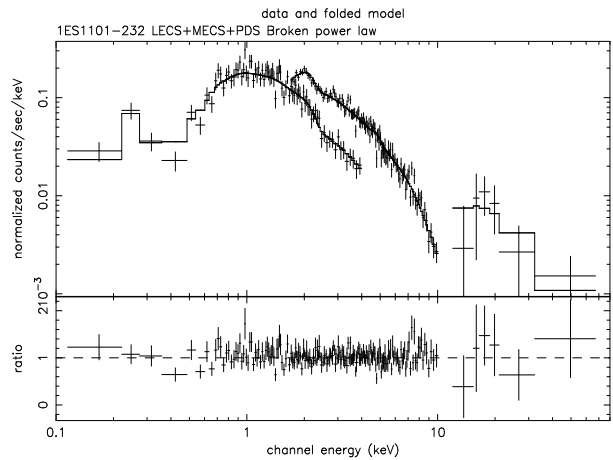}
\caption{X--ray spectrum (LECS+MECS+PDS) of June 1998, with broken power law fit
(model 3 from Table~\ref{broken_n})}
\label{fignew}
\end{figure}

The object has been observed in AO2 on 19 June 1998 for a total of 
8958 sec (LECS -- 3167 net counts); 
24895 sec (MECS(2+3) -- 10612 net counts) and 
10792 sec (PDS on source -- 1320 net counts).
The AO2 observation has been performed with only 2 MECS (MECS2 and MECS3)
since MECS1 was no longer active.
The source has been observed almost in the same period  (May 1998) by the 
Mark6 telescope (working in the GeV-TeV range). 

The extraction of the \sax\ data has been performed with FTOOLS v4.0 and the
spectral analysis with XSPEC v9.0, using the most recently available matrices
(September 1997 release).
The data analysis has been performed on the same guidelines as outlined
by the \sax\ Cookbook ({\small\verb+http://www.sdc.asi.it/software/cookbook/+})
and described e.g. in Wolter et al. (1998).
We summarize here that counts are extracted in a circular region of 8.5'/4' 
(LECS/MECS) radius, and the background is taken from the blank 
sky images distributed by the \sax\ Data Center, in a region corresponding 
to the one used to extract
source counts. Counts are binned  so as to have at least 30 total counts 
in each bin to ensure applicability of the $\chi^2$ statistics.
Fit to LECS data are performed only up to 4 keV, as the
response matrix of LECS is not well calibrated above this
energy (see Orr et al. 1998).
All confidence levels are computed using $\Delta \chi^2$ = 2.7 (corresponding
to 90\% for 1 interesting parameter), unless otherwise stated. 

For the LECS+MECS combined data, 
a single power law is not acceptable ($> 3 \sigma$) with $N_{\rm H}$
fixed at the Galactic value.
If $N_{\rm H}$ is left free, the fit with a single power law is acceptable only
at about $3 \sigma$ (Prob $\sim 1.5\%$). The residuals are however 
skewed, showing that the fit is not good.
The fit is significantly improved (F--test at $>$ 99.99 \% probability) by
using a broken power law shape. 
Results for the three models for LECS and MECS data are listed in  
Table~\ref{broken_n}.

For each observation, in the first fit 
(single power-law model with  Galactic $N_{\rm H}$) we left the 
LECS normalization free
with respect to the MECS normalization to account for the residual 
errors in intensity cross-calibration (see Cusumano, Mineo, Guainazzi
et al. in preparation). The fitted value 
of 0.696 (1997) and 0.701 (1998) fall in the range
expected given the current knowledge of the cross-calibration
(F. Fiore, private communication; see also 
{\small\verb+http://www.sdc.asi.it/software/cookbook/cross_cal.html+}).
Since the ratio of the two normalizations depends
on the position of the source in the detector, and not on the model
chosen, we fix the LECS/MECS normalization to 0.7 also for the
other subsequent models.

The PDS exposure time is not even twice than in AO1 (10.8 ks vs. 6.4 ks). 
Since the spectrum is steeper  and the source is 
fainter  than in the AO1 observation, the PDS detection is 
not more significant than the AO1 detection ($2.6 \sigma$). 
In order to fit the PDS data alone we rebin them to 6 data points. 
We fix the absorbing column to the Galactic value (N$_H = 5.76 \times 
10^{20}$ cm$^{-2}$) while the normalization with respect to the MECS
and index are left free.
The best fit slope is
$\alpha_{PDS}$ = 1.02 [$<$ 2.65], and the PDS/MECS ratio is 0.90, consistent
with what expected on the basis of cross-calibration of the instruments,
with a $\chi^2$ = 2.2(4 dof), for
a probability of 70\%, therefore statistically acceptable.
However, given the low statistical 
significance of the PDS data, the uncertainty on the slope is high. 
We therefore fit the total spectrum, from 0.1 keV to $\sim $50 keV, using 
LECS, MECS and PDS data together: the result of the broken power law fit
is shown in Figure~\ref{fignew}.  
It yields an unabsorbed flux in the [2--10 keV] band of
$F_X=2.54 \times 10^{-11}$ erg cm$^{-2}$ s$^{-1}$, 
and a corresponding luminosity in the same band of
$L_X = 4.7\times 10^{45}$  erg s$^{-1}$.
The PDS data are consistent with the LECS+MECS extrapolation.

\subsection {Comparison of the 1997 and 1998 observations}

\begin{figure}
\psfig{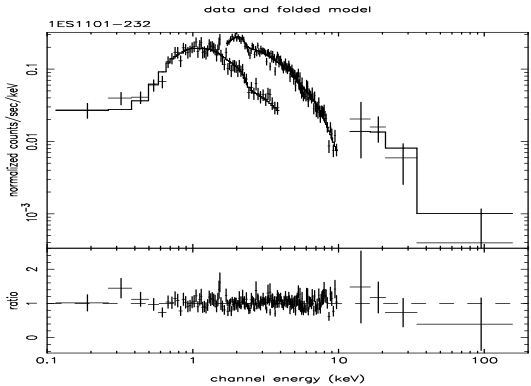}
\caption{X--ray spectrum (LECS+MECS+PDS) of Jan 1997, with broken power law fit
(model 3 from Table~\ref{broken_o}).}
\label{figold}
\end{figure}

For ease of comparison, we report in Table~\ref{broken_o} the AO1 observation 
results of 1ES~1101--232, from Wolter et al. (1998).
The LECS exposure was 5195 sec (2484 net counts), the MECS exposure
was 13830 (9509 net counts) and the PDS on-source exposure was 6410 sec (1996 
net counts).
The best fit models of the LECS+MECS spectra for various spectral 
shapes are listed in Table~\ref{broken_o}, while in Figure~\ref{figold} 
the LECS+MECS+PDS 
spectrum is plotted, with the best fit of model (3).

\begin{figure}
\psfig{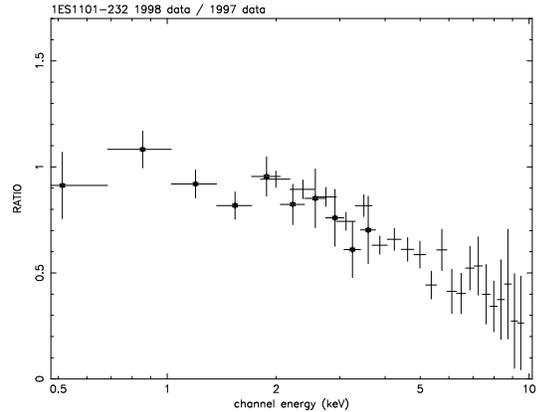}
\caption{Ratio of data point from the Jun 1998 and Jan 1997 observations. 
Stars indicate data points from the LECS instrument, crosses data points
from the MECS instrument.
The trend in the data, with a drop above $\sim$2 keV, is evident.}
\label{figrap}
\end{figure}

In order to compare the two observations, we first check directly the count
rates in the two epochs. The best representation is the ratio of
the two observed spectra, that does not depend on the choice of models
and parameters. We therefore bin the two spectra, the relative 
background and response matrices in 32 channels, in order to avoid having 
empty bins after background subtraction. We then divide the two 
background-subtracted count rates and plot the results (after having 
flagged out the energy ranges that are not well calibrated in the matrices) 
in Fig.~\ref{figrap}. 
The ratio is consistent with being flat up to $\sim 2$ keV, and steepening 
after it, showing that a change in the spectrum occurred above $\sim 2$ keV. 
The two instruments, LECS and MECS, give a consistent ratio
in the overlapping energy range. A fit with a constant in the interval   
0.5--10 keV is not statistically acceptable.

Also comparing the fit results for the two observations of Jan 1997 and 
Jun 1998, we see that the high energy slope
($\alpha_2$) is steeper in the second one. The flux, with all the
three best fit models reported in the tables, is a factor of $\sim$ 32\% 
lower in the second than in the first one. 
The low energy slope ($\alpha_1$) and the break energy ($E_0$) are
instead consistent within the errors between the two observations.

We can make therefore the hypothesis that the different fluxes and 
spectra between the two epochs are 
explained by a change in the spectral slope above the peak of the synchrotron
emission. Hence we fit all the data (1997+1998) simultaneously, keeping the low
energy slope and break energy tied (equal one to each other, but free to 
vary) between the two observations. 
On the contrary, the high energy slopes in the two observations are left 
independent.
The $N_{\rm H}$ is fixed to the galactic value, that fits well both 
observations. The PDS data are not used for the comparison, being of 
low statistical significance; the LECS and MECS data are re-binned to
100 total counts for each bin, to improve the significance of the individual
data points, since no small range feature is present. The LECS/MECS 
normalization is again fixed to 0.7.

\begin{figure}
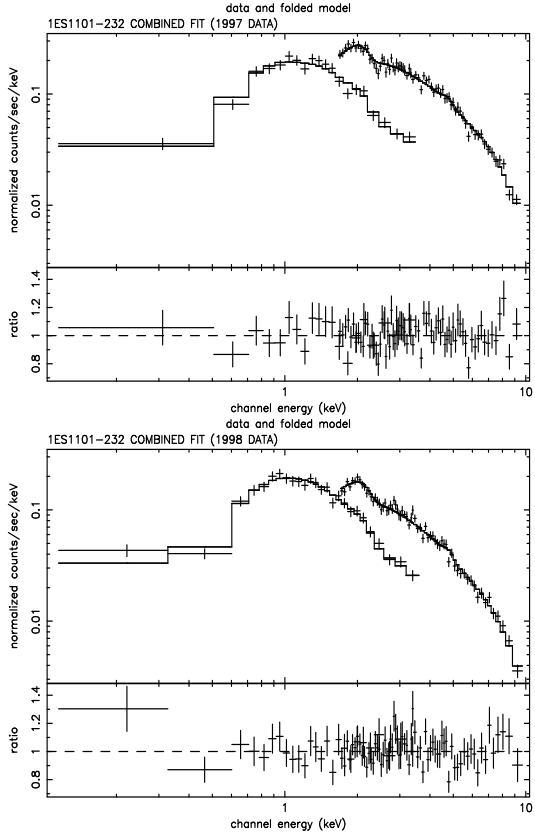

\psfig{file=9325.f4a,width=7truecm,height=5.5truecm,angle=-90}
\psfig{file=9325.f4b,width=7truecm,height=5.5truecm,angle=-90}
\caption{LECS+MECS spectrum and broken power law fit for Jan 1997 (Upper Panel)
and Jun 1998 (Lower Panel) observations. The broken power law fit of the two
observations combined is given in the text.}
\label{figcom}
\end{figure}

\begin{table}
\caption{Unabsorbed Fluxes and Rest--Frame Luminosities for the two 
observations in the (0.1--2) and (2--10) keV energy bands.}
\begin{tabular}[htb]{| l c c c c |}
            \hline
Date     & $F$[0.1-2]  & $F$[2-10]   & $L$[0.1-2] & $L$[2-10] \\
         & \multicolumn{2}{c}{$\times 10^{-11} \ecs$}&\multicolumn{2}{c|}{$\times 10^{45}$ erg s$^{-1}$} \\
\hline 
04/01/97 & 3.46 & 3.85 & 5.4 & 6.8   \\
19/06/98 & 3.32 & 2.53 & 5.3 & 4.7 \\
\hline
\end{tabular}
\label{fluxes}
\end{table}

The resulting values of $\alpha_1$ and $E_0$ are consistent with those of both
single observations: $\alpha_1$=0.72 [0.44, 0.88]; 
$E_0=1.17 [0.93, 1.43]$ keV.
The high energy slope is 
$\alpha_2^{1998}$ = 1.26 [1.19, 1.32], vs. 
$\alpha_2^{1997}$ = 1.01 [0.95, 1.07], confirming that the slope indeed
steepened significantly. The $\chi^2$ of the fit is 184.4 with 178 dof, 
corresponding to a probability of 36\%.
The errors quoted here are 90\% confidence for 
4 parameters of interest ($\Delta\chi^2 = 7.76$).
The combined spectra (LECS and MECS) for both 
observations with the best fit model are shown in Figure~\ref{figcom}.
We report in Table~\ref{fluxes} the unabsorbed fluxes and rest-frame
luminosities in the 
0.1--2 and 2--10 keV bands for the two observations using the combined 
fit results.

The fluxes derived using the combined model are consistent with the
fluxes derived from the two independent fits to the observations.  We can
therefore attribute the observed flux variation entirely to a steepening
of the high energy ($>$2 keV) portion of the spectrum.  There is no
evidence, within the statistical uncertainties, of a change in $\alpha_1$
or $E_0$, although in other well known sources (Mkn501, Mkn421, PKS2155--304) 
an increase in flux seems to be linked to an increase in $E_0$ and/or the 
high energy slope.

\section{The TeV band }

Blazars as a class have been shown to emit a large fraction of their power
at high energies, in the MeV--TeV band.
The current models assume that the high energy emission is produced by
Inverse Compton scattering (e.g. the SSC model or the EC model, see Ghisellini
et al. (1998) for the relevance of the two mechanisms), and the location of 
the peak of the Compton component depends mainly on the lower energy peak due 
to the Synchrotron component. 

In particular, objects of the LBL kind, that have the synchrotron peak
at soft energies,
show their second peak at energies in the MeV--GeV range, and are in fact 
detected by EGRET on board CGRO (see Mukherjee et al. 1997).
Objects of the HBL kind, that have their synchrotron peak at UV--X--ray frequencies,
instead, should show the Compton peak in an even higher energy band. In fact,
up to now, the only sources of VHE gamma rays (in the TeV band, by using
$\check{\rm C}$erenkov detectors) are HBL: 
Mkn 421 (Punch et al. 1992), Mkn 501 (Quinn et al. 1996), 
1ES~2344+514 (Catanese et al. 1998), 
PKS 2155--304 (Chadwick et al. 1999b).

The new $\check{\rm C}$erenkov arrays allow the detection of bright sources 
in relatively short exposure times, and therefore it has been possible to 
monitor their variability:
some of these objects have in fact shown periods of flaring activity on 
time-scales as short as 15 minutes (Gaidos et al. 1996, Aharonian et al. 1999).

Another point of debate is the amount of absorption of VHE photons due to 
the cosmic infra--red background. The aforementioned detected objects in fact
are all nearby.
Detection of sources that lie further away can therefore help in constraining
the amount of the IR background (e.g. Stecker and De Jager, 1998).

The simultaneous detection of X--ray and TeV emission allows us to estimate
a number of physical parameters of the source, such as the magnetic field and
the Doppler factor (see e.g. Tavecchio et al. 1998). 
We will show in Section 4 that a number of interesting constraints can be 
derived also by using the upper limit derived in the TeV band, together 
with the X--ray band information.

\subsection {Predictions of TeV emission in  1ES~1101--232 from phenomenological
constraints.}

A very simple prediction of the TeV emission can be made  by following 
the scheme presented e.g. in Fossati et al. (1998).
In this model a) the ratio of the frequencies of the high (Compton) and low 
(synchrotron) energy peaks is constant and equal to $5\times 10^{8}$ and
b) the high energy peak and the radio luminosity have a fixed ratio, 
$\nu_\gamma L_{peak,\gamma} \over \nu_{5GHz} L_{5GHz}$ =$3 \times 10^{3}$. 
This model represents the average SED observed for 
the various classes, while single sources can deviate from
this average phenomenological parameterization.
This very simple relationship, however, allows us to make order of magnitude
prediction even not knowing the physical conditions at the source. From 
the two expressions above we can derive both the expected frequency
of the second peak and its intensity.
The peak of the synchrotron emission is measured (this paper and Wolter 
et al. 1998) at
$\nu_s = 3 \times 10^{17}$ Hz: the Compton peak is therefore expected
at $\nu_C = 1.5 \times 10^{26}$ Hz which corresponds to $\sim$0.6 TeV.
The measured radio flux at 5 GHz is $5 \times 10^{-15}$ erg 
cm$^{-2}$ s$^{-1}$ (see references in Table~\ref{refere}).
The expected flux at the Compton peak (0.6 TeV) is therefore
$\sim 8. \times 10^{-12}$ photon cm$^{-2}$ s$^{-1}$.   

Another prediction can be made from the observed X--ray flux, by using the
recipe of Stecker, De Jager \& Salamon (1996). Within an SSC scenario they
use simple scaling arguments to predict the TeV fluxes for HBL, based on the 
X--ray flux and the assumption that the properties of the emission are similar 
to those observed for Mkn 421. Their argument is partially
supported by the actual detection of (a few) other sources
for which they predicted possible detectability.
The factor of increase between the synchrotron and Compton 
component is $10^9$, and, assuming $L_C/L_s \sim 1$,
they derive $\nu_{TeV} F_{TeV} \sim \nu_{X} F_{X}$.
Therefore, from the observed X--ray properties for 1ES~1101--232 (peak around
1 keV and flux of 2.5--3.8 $\times 10^{-11} \ecs$), we infer that the expected
Compton peak is around 1 TeV with a flux of $1.5$--$2.5 \times 10^{-11}$ 
photon cm$^{-2}$ s$^{-1}$.

These estimates make 1ES 1101--232 a good TeV candidate, at the border
of current sensitivities, and possibly well above it
during flaring activities.

\subsection{TeV Observations}

1ES~1101--232 has been observed on the nights of 19--27 May 1998
with the Durham University Mark 6 atmospheric $\check{\rm C}$erenkov telescope.
We summarize here the data analysis and the results presented in 
Chadwick et al. (1999a). 
The telescope uses the imaging technique to separate VHE gamma rays from the 
cosmic ray background, together with a robust noise-free trigger (Armstrong
et al. 1999). Data are taken in 15 minutes segments, alternating 
ON-source with an equal number of OFF-source observations. 
After removal of cloud--affected data, there are a total of
10.5 hours ON-source data and the same amount of OFF-source data.
Data are screened for ``good" events by the selection criteria listed in
Table 2 of Chadwick et al. (1999a).
The source was not detected and an upper limit of 
$F_{TeV}$ [$> 300$ GeV] = 3.7$\times 10^{-11}$ photons cm$^{-2}$ s$^{-1}$  
has been derived.

The data have been investigated for time variability on time-scales of days 
(at a flux limit of  $\sim 1 \times 10^{-10}$ cm$^{-2}$ s$^{-1}$) 
and 15 min intervals. There is no evidence for any bursting behavior
(Chadwick, private communication).

\section {SED construction and TeV predictions from theoretical models.}

\begin{figure}
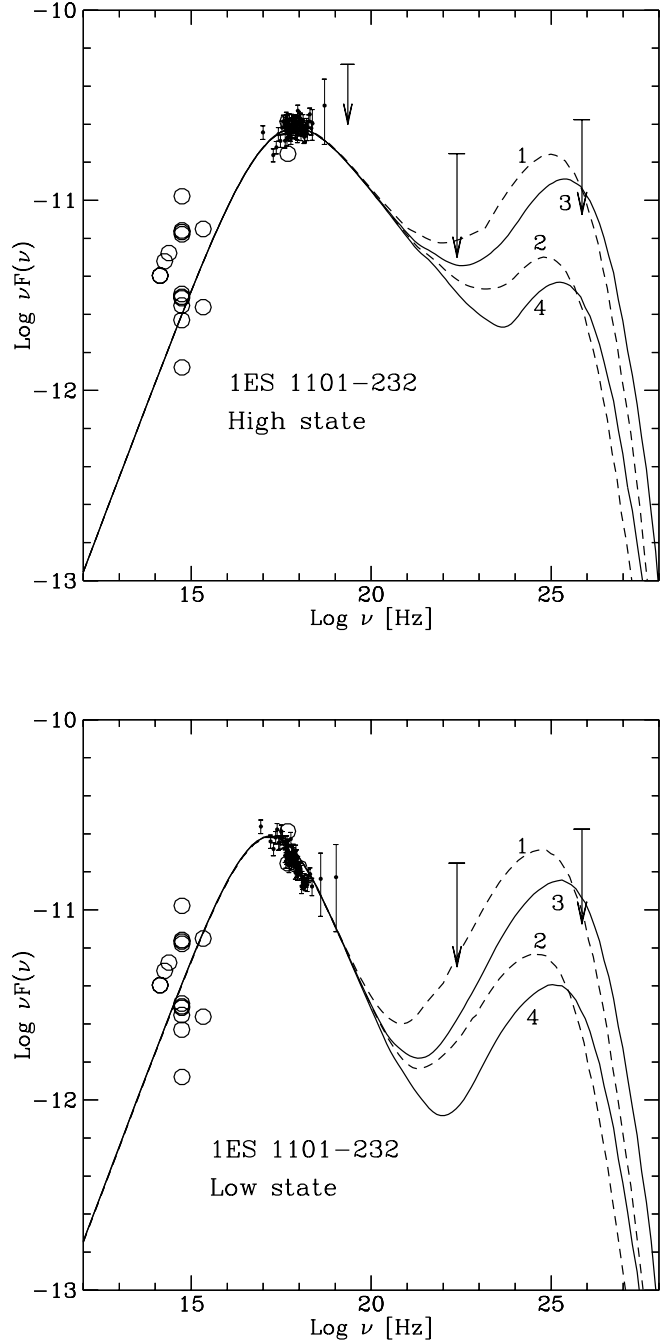

\psfig{file=9325.f5a,width=9.5truecm}
\psfig{file=9325.f5b,width=9.5truecm}
\caption{SED of the two \sax\ epoch observations {\it Top:} 04/01/97 -- 
high state; {\it Bottom:} 19/06/98 -- low state, plus models.
Large empty circles represent historical data from literature, while arrows
indicate upper limits from EGRET (MeV) and Mark6 (GeV-TeV) experiments
respectively
(see Table~\ref{refere} for references). Small dots with error-bars are 
the unfolded X--ray spectra discussed in this work.
Lines represent the spectra computed with the SSC model (see text for
details) with different model parameters, reported in Table~\ref{params}:
dashed lines represent spectra calculated with $\delta=10$ and two different 
values for the magnetic field $B$; similarly, solid lines represent spectra 
calculated with $\delta=20$ and two different values for the magnetic 
field $B$. 
}
\label{figsedb}
\end{figure}

We have constructed the overall Spectral Energy Distribution (SED), using
both data from literature and the \sax\ spectra from the two observations. 
We construct two different SED, one for each \sax\ observation, that are
clearly modeled by a different synchrotron state.
The two SED are presented in Fig.~\ref{figsedb}.

\begin{table}
\caption{References for the data points of Fig~\ref{figsedb}.}
\begin{tabular}[h]{|l l|}
\hline
Band      	& Reference	\\
\hline
Radio 1.4GHz    & NVSS: Condon et al. 1998 AJ 115, 1693  \\
Radio 1.4GHz    & Remillard et al. 1989 ApJ 345, 140 \\
Radio 5 GHz     & Perlman et al.  1996 ApJS 104, 251 \\
Radio 5 GHz     & Giommi et al.     1995 A\&AS 109, 267 \\
Optical V 	& Pesce et al.   1994 AJ 107, 494 \\
Optical V       & Lanzetta et al. 1995 ApJ 440, 435 \\
Optical V       & Giommi et al.       1995 A\&AS 109, 267 \\
Optical V       & Jannuzi et al. 1994 ApJ 428, 130 \\
Optical V       & Falomo et al.       1994 ApJS 93, 125 \\
IR J,H,K        & Bersanelli et al.   1992 AJ 104, 28 \\
IR K		& Falomo et al.       1993 AJ 106, 11 \\
UV 1400A  	& Pian \& Treves    1993 ApJ 416, 130 \\
UV 1400A  	& Edelson et al.   1992 ApJS 83, 1 \\
X--ray 2 keV     & Perlman et al. 1996 ApJS 104, 251 \\
X--ray 2 keV     & Giommi et al.    1995 A\&AS 109, 267  \\
EGRET 		& Fichtel et al. 1994 ApJS 94, 551  \\
\hline
\end{tabular}
\label{refere}
\end{table}

\begin{table}
\caption{Input parameters for the models of Fig.~\ref{figsedb}}

\begin{tabular}[h]{| r r c c c c c c|}
\hline
\multicolumn{3}{|l}{High State} & & & & & \\
\hline
 \#   &$\delta$ &$B$ &$\gamma_b$  &$R$     &$K$           &$n_1$ &$n_2$ \\
      &         &(G) &$\times10^4$ &cm     &cm$^{-3}$     &      & \\
\hline
1 & 10  & 0.3 & 15.8 &1$\times10^{16}$ & 1.7$\times10^4$  & 2 & 3.5\\
2 & 10  & 0.6 & 11.2 &1$\times10^{16}$ & 5.9$\times10^4$  & 2 & 3.5\\
3 & 20  & 0.1 & 19.3 &1$\times10^{16}$ & 7.9$\times10^3$  & 2 & 3.5\\
4 & 20  & 0.2 & 13.5 &1$\times10^{16}$ & 2.8$\times10^4$  & 2 & 3.5\\
\hline
\hline
\multicolumn{3}{|l}{Low State} & & & & & \\
\hline
 \#   &$\delta$ &$B$ &$\gamma_b$  &$R$     &$K$           &$n_1$ &$n_2$ \\
      &         &(G) &$\times10^4$ &cm     &cm$^{-3}$     &      & \\
\hline
1 & 10  & 0.5 & 8.45 & 1$\times10^{16}$ &  2.5$\times10^4$ & 2  & 3.9\\
2 & 10  & 1.0 & 5.98 & 1$\times10^{16}$ &  8.9$\times10^3$ & 2  & 3.9\\
3 & 20  & 0.1 & 13.3 & 1$\times10^{16}$ &  6.3$\times10^3$ & 2  & 3.9\\
4 & 20  & 0.2 & 9.40 & 1$\times10^{16}$ &  2.2$\times10^3$ & 2  & 3.9\\
\hline
\end{tabular}
\label{params}
\end{table}

We have reproduced the SED in both X--ray states with a simple homogeneous SSC 
model.
The source is modeled as a spherical region with size $R$, uniform and
tangled magnetic field $B$, in motion toward the observer with a bulk 
Lorentz factor $\Gamma$.
The region is filled by a population of relativistic electrons with a
distribution of Lorentz factors given by:
$N(\gamma )=K\gamma^{-n_1}(1+\gamma/\gamma_{b})^{n_1-n_2}$, 
where the asymptotic slopes are $n_1$ and $n_2$, the break point
is $\gamma_{b}$ and $K$ is a normalization factor.
The self--Compton emission is derived taking into account the full 
Klein--Nishina (KN) cross section, computed using the relations reported in 
Jones (1968; see also Blumenthal \& Gould 1970). 
We do not take into account absorption of IR photons by the infrared
background, whose emission level is still uncertain. 
This implies that in principle the derived curves are upper limits to the
detectable VHE emission, also because the redshift of 1ES~1101--232 is only
slightly smaller
than the limit ($z=0.2$) chosen to monitor blazar emission in the TeV band
by the HEGRA experiment (Rhode \& Meyer, 1997).
On the other hand, a detection of the source in this band would provide also a
measure of the density of IR background photons.

Although the complete determination of the set of physical parameters for the
SSC model requires the knowledge of the positions of both the synchrotron peak 
and the Inverse Compton peak (as discussed in Tavecchio et al. 1998), 
we can put strong constraints to the parameters using the informations
provided by the X--ray spectrum and the TeV upper limit
suggesting that the condition $L_C/L_s\leq 1$ applies.

We note here that the analytical relations discussed in Tavecchio et al. (1998)
in the KN regime are obtained with a step approximation for the KN 
cross--section. 
In the extreme KN regime the numerical values given by this 
approximation might differ from the results of the numerical model derived
with the full KN treatment, but we can use the analytical discussion as a 
guideline for the numerical model.

Typical variability time-scales observed in HBLs 
($t_{var}\sim\,10^{3-4}$ s, see e.g
Zhang et al. 2000, Giommi et al. 1999, 2000) suggest Doppler factors 
$\delta$ in the range 10--20 and sizes of $R \leq 10^{16}$ cm. 
We fixed the radius to $R=10^{16}$ cm. This choice directly puts a lower 
limit to the value of the magnetic field:

\footnotesize
\begin{equation}
B\delta^{2+\alpha _1} >  A \times
 (1+z)^{\alpha _1} \left[ \frac{g} 
{\nu_c\nu_s} \right] ^{(1-\alpha _1)/2} 
\left( \frac {L_s}
{  t_{var} L_C^{1/2}} \right)
\label{ulower} 
\end{equation} 
\normalsize

\noindent
where $g(\alpha _1,\alpha _2)$ is a constant related to
the spectral indices $\alpha_1$ and $\alpha _2$ and A is the
appropriate constant
(see Tavecchio et al. 1998 for details on (\ref{ulower})).
From this relation we can infer that an upper limit
on $L_C$ and an estimate of $t_{var}$ directly puts a lower 
limit for $B$. The physical reason for this lower limit is related to the 
fact that the synchrotron peak Luminosity is proportional to the product
$N_eB^2$ (where $N_e$ is the number of emitting electrons), while the ratio
between the Compton peak Luminosity and the synchrotron peak Luminosity is 
proportional to $N_e$: therefore an upper limit to $L_C/L_s$ gives an upper 
limit to $N_e$ and this, together with the synchrotron Luminosity, provides 
the lower limit for the magnetic field $B$.
Another way of describing the effect, assuming that
the scattering is in the Thomson regime, is that in this case
requiring $L_C/L_s<1$ corresponds to require that the
ratio between the synchrotron and the magnetic energy densities 
$U_{syn}/U_B<1$.
For a given source size and Doppler factor the synchrotron radiation 
energy density is fixed, and the above relation then corresponds to a 
lower limit on the value of the magnetic field.

In Fig.~\ref{figsedb} we plot the spectrum of both states computed for 
two different values of $\delta$ ($\delta=10$ for the dashed lines,
$\delta=20$ for the solid lines), and for two different values of the magnetic
field $B$, as listed in Table~\ref{params}.
The lowest value of $B$ has been determined by requiring
not to over-produce the high energy (TeV) emission, since, for a given
synchrotron luminosity, size and Doppler factor, the ratio
between the self Compton and the synchrotron powers depends only on the
magnetic field.

We take this value, listed in Table~\ref{params},
for the models 1 and 3 shown in Fig. 5.
For models 2 and 4 we have doubled the B value, and decreased the
relativistic electron density and $\gamma_b$ accordingly, in order to produce
the same amount of synchrotron flux and about the same synchrotron peak
frequency.
In this case the self--Compton flux decreases, due to the decreased
electron density.

The transition from the high to the low state is consistent with a change
of the second slope $n_2$ and with a decrease of $\gamma_b$ by a factor
of 1.5--2.  This behavior is similar to what observed in the other well
known TeV BL Lac, such as PKS 2155--304 (see e.g. Chiappetti et al. 1999),
Mkn 501 (e.g. Pian et al 1998) and Mkn 421 (Maraschi et al. 1999), where
high X-ray states are interpreted as states with either higher 
$\gamma _b$ and/or higher magnetic field.

It is evident from Fig.~\ref{figsedb} that a small change in the magnetic
field, while still consistent with the X--ray (\sax) observations,
produces a dramatically different amount of TeV photons. Assuming that
the size of the source and the Doppler factor do not vary substantially
with time, variations of the synchrotron flux can be attributed to
changes of the density of electrons and/or the magnetic field.  If this
is the case, we expect that the TeV emission can be easily detected,
either for X--ray fluxes slightly brighter than what observed up to now,
or by longer TeV exposure times.

\section{Results and Conclusions}
   
The X--ray spectrum of 1ES~1101--232 as measured by \sax\ is fitted 
only by a broken power law (a single power law or an absorbed 
power law are not statistically acceptable)
with a break at 1.3 - 1.9 keV. From the first to the second observation, 
the spectrum varied at high energies, becoming softer (steeper). 
The flux decrease, by about 32\%, has occurred in the 2--10 keV band.
The PDS observation are not of statistical significance sufficient
to put a real constraint on the spectrum.

Even if the variation in the X--ray band is not dramatic, we can clearly
distinguish between the two states, that are modeled by different parameters
of the synchrotron component.

By using the TeV upper limit and the two \sax\ observations we model also the
higher energy portion of the spectrum as a self--Compton component, by
using the model described e.g. in Tavecchio et al. (1998) that assumes a simple
homogeneous SSC model in the KN regime, in which the relativistic electrons
have a broken power law energy distribution. 

The two X-ray states of the source are described by varying this
distribution, assuming that the other relevant parameters ($R$ and
$\delta$) are nearly constant.

We can compare these results with what found for the few TeV detected sources.
The choice of Doppler factor of 10 and 20 made here is in the interval of
the values of $\delta$ found by other authors for Mkn 421, 
PKS2155--304 and Mkn501, that range
from $\delta \sim 5$ (e.g. Takahashi, 1999; Mkn 421, Catanese et al. 
1998: Mkn 501) 
to $\delta \sim 30$ (e.g. Bednarek \& Protheroe 1999: Mkn501; 
Kataoka et al 2000: PKS2155--304;
Bednarek \& Protheroe 1997: Mkn 421).
At the same time, values of $B$ in excess of 0.03 G and up to 1 G 
(Chiappetti et al. 1999) are found for the same sources.

Of the three above mentioned objects, the most similar to 1ES~1101--232 
is Mkn 501, for which different measures have been produced: e.g.
$\delta \geq 15$ \& $B=0.8 $G (Pian et al. 1998);
$\delta = 15$ \& $B=0.2 $G (Kataoka et al. 1999);
$\delta = 30$ \& $B=0.7 $G (Bednarek \& Protheroe 1999),
in good agreement with the values of Table~\ref{params}.

Even if based, besides the accurate X-ray spectral determination up 
to $\sim 50$ keV, only on a TeV upper limit we can infer that
the physical conditions in 1ES~1101--232 are similar to the
brightest TeV sources, making it a very promising candidate for 
TeV observations, and a testbed for the SSC model.
Furthermore, since the redshift of 1ES~1101--232 is intermediate,
a detection of this source in the VHE range would pose constraints
on the density of the IR background photons that is still at the moment
very uncertain.

\begin{acknowledgements}
This work has received partial financial support from the Italian 
Space Agency. 
We would like to thank Paula Chadwick and S.J. McQueen for informing us
about their VHE results in advance of publication, Paolo Giommi and Roberto
Della Ceca for helpful discussion and comments, and an anonymous referee
for useful suggestions that improved the readability of the paper.

\end{acknowledgements}

\end{document}